\documentclass[onecolumn]{IEEEtran}
\usepackage[utf8]{inputenc}
\usepackage[normalem]{ulem}
\usepackage{amsmath}
\usepackage{amsfonts}
\usepackage{amssymb}
\usepackage{algpseudocode}
\usepackage{algorithm}
\usepackage[pdftex]{graphicx}
\usepackage{epstopdf,epsfig}
\usepackage{enumerate}
\usepackage{pgfplots}
\usepackage{bbold}
\usepackage{subfigure}
\usepackage{mathtools}
\usepackage{amsthm}
\usepackage{wrapfig}
\usepackage{capt-of}
\usepackage[english]{babel}
\usepackage{url}
\usepackage{color}

\ifCLASSOPTIONcompsoc
  \usepackage[nocompress]{cite}
\else
  \usepackage{cite}
\fi

\theoremstyle{remark}

\theoremstyle{theorem}

\theoremstyle{definition}

\title{Using mm-Waves for Secret Key Establishment}

\author{
\IEEEauthorblockN{Mohammed Karmoose$^1$, Christina Fragouli$^1$, Suhas Diggavi$^1$, Rafael Misoczki$^2$, Lily L. Yang$^2$, Zhenliang Zhang$^3$
}\thanks{\noindent$^1$ The authors are with the Electrical and Computer Engineering department, University of California at Los Angeles, CA 90055 (e-mail: mkarmoose@ucla.edu, christina.fragouli@ucla.edu, suhasdiggavi@ucla.edu).}
\thanks{$^2$ The authors are with the Security and Privacy Research Group, Intel Labs (e-mail: rafael.misoczki@intel.com, lily.l.yang@intel.com).}
\thanks{$^3$ The author is Alibaba.com (e-mail: zzl.csu@gmail.com).}
\thanks{A short version of this manuscript is in the IEEE Communication Letters~\cite{8684919}.}
}

\begin{document}

\maketitle
\begin{abstract}
The fact that Millimeter Wave (mmWave) communication needs to be directional is usually perceived as a challenge; in this paper we argue that it enables {efficient secret key sharing} that are unconditionally secure from passive eavesdroppers, by building on packet erasures. We showcase the potential of our approach in two setups: mmWave-based WiFi networks and vehicle platooning. 
We show that in the first case, we can establish a few hundred secret bits with minimal changes to standard communication protocol; 
while in both cases, with the right choice of parameters, we can potentially establish keys in the order of tenths of Mbps. These first results are based on some simplifying assumptions, yet we believe they give incentives to further explore such techniques.
\end{abstract}

\section{Introduction}

Millimeter Wave (mmWave) communications are expected to have significant impact on wireless communication networks such as 5G networks~\cite{wang2014cellular}, over-$60$ GHz-WiFi networks~\cite{ghasempour2017ieee}, autonomous vehicles and vehicle platoons~\cite{ploeg2011design}. In addition, the inherent directionality of mmWaves can be utilized to establish physical layer secrecy. In fact, inherent properties to the mmWave communication systems (e.g., channel variability, directionality, wider bandwidth allocations) are exploited to reduce the eavesdropping capabilities of Eve~\cite{zhu2017secure}.

This work proposes a secret key establishment technique based on mmWave communication. The main motivation stems from the fact that packet erasures can help create secrecy. In fact, a recent line of work~\cite{argyraki2013creating,safaka2016creating} demonstrated secret key establishment protocols that relied on packet erasures from multi-hop and multipath communication, as well as the use of wiretap codes and beamforming over WiFi. Experiments showed that these protocols yield several Mbps of shared secret keys. This work extends these secret key exchange protocols in a mmWave environment. In fact, the mmWave directional transmissions, if not perfectly aligned, inherently lead to packet losses, and thus it seems a natural host environment for erasure-based key {establishment}. 
Differently from existing physical layer secrecy works~\cite{hamamreh2018classifications}, our proposed protocol: 1) is information-theoretically secure against passive eavesdroppers with limited network presence, 2) allows for very high secret key exchange rates, and 3) is autonomous ({\it i.e.}, does not require third-party assistance). 

Our proposed technique differs from existing cryptographic encryption measures which depend on limited adversarial capabilities: computational capabilities, {\it e.g.} Diffie-Hellman (DH), or storage capabilities, {\it e.g.}, Bounded-Storage Model~\cite{dziembowski2006intrusion}. 
In contrast, our proposed scheme establishes secret keys that are information-theoretically secure against eavesdroppers with limited network presence. Moreover, current cryptographic techniques rely on high complexity algorithms to compensate for the low rate of secret key establishment. In this work, we show, through two different scenarios, that our scheme promises secret key generation rates in the order of tens/hundreds of Mbps. \\
\noindent\textbf{Main Contributions.} We showcase our approach for two scenarios: (1)  over-$60$ GHz-WiFi networks, where base stations use mmWave antenna arrays for transmissions. First, we propose an analytical model for the instantaneous received Signal-to-Noise-Ratio (SNR), that is inspired from the empirical channel model in~\cite{thomas20143d} and system parameters ({\it e.g.}, antenna array sizes and beam patterns as described in~\cite{ghasempour2017ieee}.
We show that, with the right choice of parameters,
with minor modification to the standard beamsweeping mechanism, a considerable amount of secret bits (up to {a few hundreds}) can be established between the base station and mobile devices for virtually no additional transmission cost. 
In addition, we show that a more invasive secret key establishment protocol achieves few hundred Mbps of secret key generation rates with increased security guarantees.\\
(2) Vehicular platooning, which is a safety-critical application. We show that,  with appropriate choices of code parameters and antenna placement, our technique allows platoons to establish keys with rates up to {$166$ Mbps} -- $4$ orders of magnitude gain over rates achieved by DH; this allows the use of (otherwise impractical) One-Time Pad (OTP) encryption (an information-theoretically secure encryption technique).

The paper is organized as follows: Section~\ref{Sec:model} presents our adversary model and background; Section~\ref{Sec:beam} {discusses the WiFi network} application; Section~\ref{Sec:platoon} {discusses} the vehicle platooning application and Section~\ref{Sec:concl} concludes the paper.

\section{Model and Background}
\label{Sec:model}

\noindent{\bf System and Adversary Model.} We consider a pair of communicating parties who wish to establish a pairwise key using the scheme in~\cite{argyraki2013creating,safaka2016creating}. The transmitting, {\it a.k.a.} Alice (resp. receiving, {\it a.k.a.} Bob) party is connected to a set of $N_T$ transmitting (resp. $N_R$ receiving) mmWave antenna arrays, each labeled by $t_i$ and situated at location $\mathcal{T}_i, i \in \{1,\cdots,N_T \}$ (resp. $r_j$ and $\mathcal{R}_j, j \in \{1,\cdots,N_R \}$). 
In addition, the communicating parties wish to communicate secretly in the presence of an eavesdropper ({\it a.k.a.} Eve), which is equipped with a set of $N_E$ antenna arrays, each label by $e_k$ and situated at location $\mathcal{E}_k, k \in \{1,\cdots,N_E \}$.
We assume Eve to be located anywhere within the transmission radius of the communicating parties, and is passive; therefore the locations of its antennas are unknown.
We assume also that Eve has access to the same physical layer technology as the legitimate nodes, has infinite memory as well as unbounded computational capabilities at her disposal, and has perfect knowledge of the protocols. 
The transmitting power used by each transmitting antenna is denoted by $P_T$, while the noise figure at each receiving antenna array (for both Bob and Eve) is $N_o$.
We assume that the available bandwidth is $B$.
We assume that each transmitting antenna array is capable of focusing its transmitting energy in desired directions by the use of appropriate beamforming mechanisms.
Given a particular beamforming direction, the received SNR at the $j$th receiver (resp. $k$th eavesdropper) antenna from the $i$th transmitter antenna is denoted by $\gamma^{(t_i,r_j)}$ (resp. $\gamma^{(t_i,e_k)}$).
Considering the fact that the wireless channel is typically random, then $\gamma^{(t_i,r_j)}$ and $\gamma^{(t_i,e_k)}$ are considered as random variables, with distributions denoted as $f_{\gamma^{(t_i,r_j)}}$ and $f_{\gamma^{(t_i,e_k)}}$.

\noindent{\bf mmWave Channel Model and Antenna Patterns.}
In mmWaves, transmitters are expected to employ transmit beamforming in order to focus transmission energy in a particular direction in space.
However, the radiated energy pattern in space as a result of beamforming strongly relies on 1) the wireless channel between the transmitters and receivers, and 2) the assumed antenna radiation pattern. Therefore, in this work, we strive to employ realistic channel models and antenna patterns in order to give a realistic assessment of our proposed mechanisms. In particular,
 (1) For over-$60$ GHz-WiFi cellular networks, we implement the point-to-point 73 GHz outdoor channel model proposed in~\cite{thomas20143d} which takes into account line-of-sight as well as multipath fading signal components. Moreover, in order to take into account the fact that transmitters/receivers that are close by in space exhibit similar channel characteristic, we also implement space consistency between receivers and transmitters, as specified in~\cite{3gppChannel}. We also use the standardized antenna radiation pattern proposed in~\cite{3gppChannel}. 
 Based on empirical data, we deduce an analytical expression for the received SNR which we describe in the next section.\\
  (2) For vehicular networks, {similar models for mmWave channel models are lacking.}
We developed instead a channel model based on ray tracing, which takes into account reflections off the hood, back and roof of the cars in the platoon. We also used a realistic model for a vertically-polarized 70 GHz antenna array system.

\noindent{\bf Secret Key Protocol \cite{czap2015secret}.}
The protocol proposed in~\cite{czap2015secret} allows Alice and Bob (each with possibly multiple antennas) to establish a shared key which is secret from Eve.
We here briefly explain how the protocol works, and delegate the details of the protocol to~\cite{czap2015secret,argyraki2013creating}.
The protocol operates in two rounds of transmission. \\
{\noindent\underline{Round 1:} Alice sends a set of random packets to Bob, who sends feedback of which packets were correctly received.
The key idea is that some of these packets are received by Bob while being erased for Eve.
If Alice knows a {\it lower-bound estimate $N$} of such erased packets, then it can create a shared secret key of size $N$ with Bob in the second round.\\
{\noindent\underline{Round 2:} Assume that $x_1, \cdots, x_M$ are the $M$ packets now shared between Alice and Bob, and Eve knows $M-N$ of them. Then, the secure common key is the concatenation of $y_1, \cdots, y_N$, where the packets $y_i$ are carefully designed (based on MDS codes~\cite{safaka2016creating}) linear combinations of $x_1,\cdots,x_M$.}
Note that Alice does not need to know which $N$ packets are erased; only the number of such packets suffices.

\noindent\textit{Example.} In round 1, Alice sends $x_1$ to $x_4$ to Bob, who sends a feedback that packets $x_1$ to $x_3$ were correctly received. Assuming Eve missed two packets $N=2$, then in round 2, the secret key is the concatenation of $y_1 = x_1 + x_2$ and $y_2 = x_2 + x_3$ which Eve would know nothing about.
}

Note that this secret is created by knowing that Eve misses at least two packets, but not necessarily knowing {\it which} two exactly.
The security is guaranteed by the fact that  the second round does not involve sending the secret itself but rather the packet indices used to create the secrets~\cite{safaka2016creating} ({\it e.g.}, the indices $(1,2)$ and $(2,3)$ in the discussed example).
In our setting, we make worst-case estimates on how much Eve misses {{\it i.e.}, $N$}, and assess how good these estimates are via the \textit{insecure areas} concept as we show next.

\begin{figure*}
 \centering
 \begin{minipage}{0.15\textwidth}
  \includegraphics[width=\columnwidth]{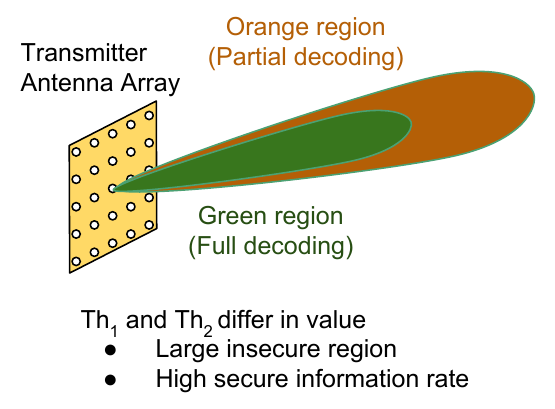}
 \caption{Transmission with wiretap codes - Code parameters values largely differ.}
 \label{fig::WiretapCodeParamChoice}
 \end{minipage}
\begin{minipage}{0.15\textwidth}
  \includegraphics[width=\columnwidth]{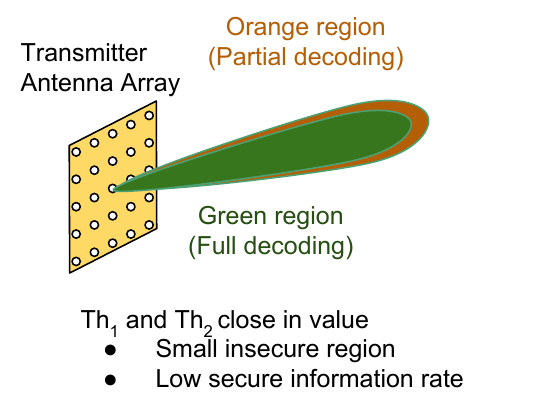}
 \caption{Transmission with wiretap codes - Code parameters are close in value.}
 \label{fig::WiretapCodeParamChoice2}
 \end{minipage}
 \begin{minipage}{0.3\textwidth}
  \centering
 \includegraphics[width=\columnwidth]{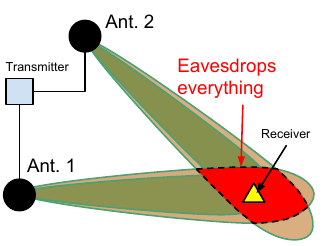}
 \caption{An illustrative example of the procotol.}
 \label{fig::ProtocolSetup}
 \end{minipage}
 \begin{minipage}{0.3\textwidth}
  \centering
  \includegraphics[width=\columnwidth]{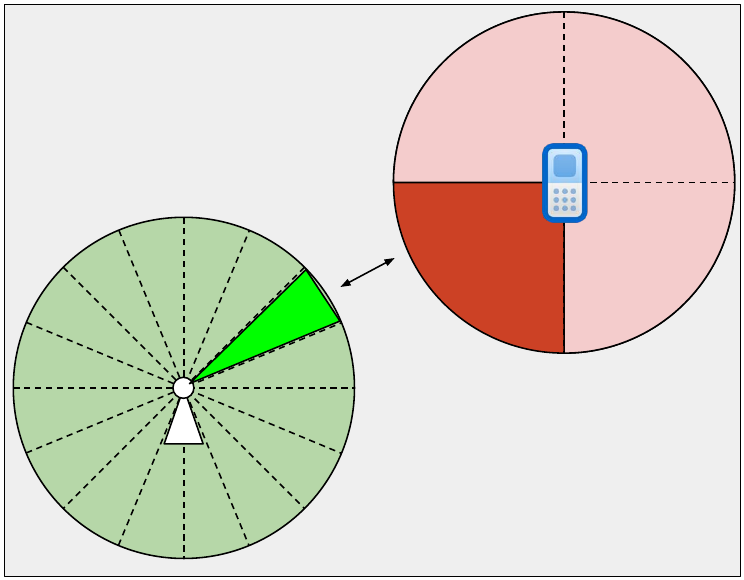}
  \caption{Antenna sectors~\cite{ghasempour2017ieee}.}
  \label{fig::AntSect}
 \end{minipage}
\end{figure*}

\noindent{\bf Creating Erasures and Insecure Areas.} The method we follow to enforce erasures in the protocol described previously is based on wiretap codes and directionality. A high-level description of wiretap codes is as follows. The performance of a wiretap code is dictated through two parameters, namely $\text{Th}_1$ and $\text{Th}_2$. 
Three distinct situations can occur when a wiretap-coded packet is received: 1) if it is received with $\gamma \geq \text{Th}_1$ then it is decoded perfectly ({\it i.e.}, ``received''), 2) if $\gamma \leq \text{Th}_2$ then it is completely missed by the receiver ({\it i.e.}, ``erased''), and 3) if $\text{Th}_1 > \gamma > \text{Th}_2$ then partial information can be extracted from the packet.
The three aforementioned modes of reception are shown in Figure \ref{fig::ProtocolSetup}. The green area highlights an area in space where a receiver would experience a value of $\gamma \geq \text{Th}_1$ and therefore would decode all transmitted information. The orange region (which typically encloses the green region) highlights an area where  $\text{Th}_1 > \gamma > \text{Th}_2$ and therefore a receiver may decode part of the transmitted information. Finally, a receiver outside the green and orange regions (white region) will not be able to infer any information.
We assume that packets transmitted in the first round are encoded using wiretap codes. Therefore, Alice hopes that Bob receives these packets while at least some of the packets are erased at Eve; we refer to this event as the {\it Secret Reception (SecRec) event}. 
In our setup, we always make the assumption that $N=1$, {\it i.e.}, packets from {\it at least} one transmission/receiver link are erased at Eve while received by Bob.
More formally, we denote by a SecRec event that there exists $(t_i,r_j)$ for $(i,j) \in \{1,\cdots,T \} \times \{1,\cdots,R \}$ for which the transmission is correctly received by $r_j$ and is erased by {\it all} eavesdropper's antennas. The probability of such event is defined as

 \begin{align}
 P_{\text{SecRec}} &\coloneqq \text{Pr}\left( \exists (i,j) : \gamma^{(t_i,r_j)} \geq \text{Th}_1, \gamma^{(t_i,e_k)} \leq \text{Th}_2 \: ; \: \forall k \right) \label{eq::prob_secrec} 
\end{align}

If $M$ is the number of packets sent on one such link, then the protocol can create shared secret key of size $N = M \cdot L$ where $L$ is the size of one packet. 
Note that increasing the number of transmitters/receivers (and dispersing them geographically) increase the probability of this estimate of $N$.
The protocol fails when the Secure Reception event does not happen, \textit{i.e.}, either Bob does not receive data or Eve receives \text{all} packets the Bob does.
Consider the example in Figure~\ref{fig::ProtocolSetup} with two transmitting antennas at Alice, and assume that Bob and Eve each has one antenna. If Eve resides in the ``not white'' region of any link then it would be able to receive the packets transmitted on that link; Secure Reception will not occur with high probability.
 We therefore consider the protocol to be vulnerable if $P_{\text{SecRec}}$ is not high enough, {\it i.e.}, $P_{\text{SecRec}} \leq 1 - \delta$ where $\delta$ is our security level. We refer to this situation as the protocol being not $\delta$-secure.
The region in space where this occurs {({\it i.e.}, the probability is not high enough)} is referred to as the \textit{Insecure Area (IA)}. Other mechanisms may be needed to protect against eavesdroppers in the IA, and therefore a smaller IA indicates a stronger key agreement mechanism. 
The choices of $\text{Th}_1$ and $\text{Th}_2$ affect the secret key generation rate as well as the size of the IA. In what follows, we use these two quantities as the performance metrics of our proposed protocol.

\noindent{\bf {Performance Metrics.}} 
{We define the following two performance metrics:
\begin{enumerate}
 \item {\it The average secret key rate:} the average number of bits per second established between the communicating parties secretly from the eavesdropper. Given a wiretap code with parameters $\text{Th}_1$ and $\text{Th}_2$, a key generation rate equal to $B\left[ \log(1 + \text{Th}_1) - \log(1 + \text{Th}_2)\right]$ can be established between communicating parties while being secure from an eavesdropper, assuming that a secret reception event occurs. Therefore, the average secret key rate is equal to 
 
 \begin{align}
R_{\text{av}} &= R_{\max}  P_{\text{SecRec}}, & R_{\max} = \left[\underbrace{B\log_2(1+\text{Th}_1)}_{\text{Decoding Rate}} - \underbrace{B\log_2(1+\text{Th}_2)}_{\text{Secrecy Overhead}}\right]. \label{eq::sec_key_gen_rate}
 \end{align}
 
 The Decoding Rate component corresponds to the raw data transmission rate achieved between the $(i,j)$-th transmitting/receiving antennas whenever $\gamma^{(t_i,r_j)} \geq \text{Th}_1$. The Secrecy Overhead component accounts for the coding overhead due to the use of wiretap codes, thus the difference is the achieved data throughput.
%
 
 \item \it $\delta$-Insecure Area or $IA_{\delta}$:
 we define the set $\mathcal{A} = \left\{ r,\theta : P_{\text{SecRec}} \leq 1-\delta \right\}$ as the set of locations in space where the protocol is not $\delta$-secure, where $r,\theta$ are polar coordinates. Therefore, we can define the $\delta$-Insecure Area as 
 \begin{align*}
IA_\delta \coloneqq \int_{r,\theta \in \mathcal{A}}  r dr d\theta.
 \end{align*}
$\delta$-Insecure Area captures the regions where the likelihood of Eve breaking the secret key establishment mechanism is too high ({\it i.e.}, at least $\delta$).
\end{enumerate}

Choosing $\text{Th}_1$ and $\text{Th}_2$ gives contradicting effects with respect to the last two objectives.
Specifically, when $\text{Th}_1$ and $\text{Th}_2$ are relatively different in value, this results in a relatively larger $P_{\text{SecRec}}$ (therefore a larger insecure area) and larger value of $R_{\max}$. The reverse effect happens when $\text{Th}_1$ and $\text{Th}_2$ are relatively close.
We finally note that today a number of practical designs for wiretap codes are emerging, based on polar~\cite{mahdavifar2011achieving}, LDPC~\cite{thangaraj2007applications} and lattice codes~\cite{lu2014usrp}, which enable with low complexity to achieve performance curves similar to $R_{\max}$ in~\eqref{eq::sec_key_gen_rate}. For this paper, we will directly use the expression in~\eqref{eq::sec_key_gen_rate} to estimate potential benefits and trade-offs.

\section{Showcase I - IEEE 802.11ay}
\label{Sec:beam}
Our first showcase application is in the context of 60-GHz-based WiFi networks \cite{ghasempour2017ieee}. The IEEE 802.11ay amendment proposes the use of directional communication to cope with the increased signal attenuation that accompanies transmission in the mmWave band.

\noindent\textbf{Directional Communication.} IEEE 802.11ay proposes the use of \textit{virtual antenna sectors} which discretizes the azimuth angle. Shown in Figure \ref{fig::AntSect}, a base station sectorizes the azimuth range into $32-64$ sectors. Being equipped with up to $3$ antenna arrays, each array is responsible from transmission in one-third of these sectors\footnote{Antenna arrays do not cooperatively transmit in the same sector.}. A mobile device is typically equipped with one antenna array and can have up to $4$ sectors. Each device has a set of pre-computed beamforming weights that correspond to transmission in each of the predefined sectors. When a base station wishes to communicate with a mobile device, both communicating parties have to agree on the best sector to use (i.e. best set of beamforming weights to employ) so that received signal strength is maximized. This sector training phase is referred to as \textit{the beamsweeping phase}, and it is split into to sub-phases: 1) a \textit{Sector-Level Sweep (SLS)} phase where both communicating parties agree on the best two sectors to use, and 2) a \textit{Beam Refinement Phase (BRP)} in which the predefined beamforming weights are fine-tuned to further maximize the received signal strength. The SLS phase is also comprised of two sub-phases: the Transmit-SLS for negotiating the best sector to use at the transmitter, and the Receive-SLS for the receiver. We claim that the proposed mechanism for beam training in IEEE 802.11ay creates an excellent opportunity to establish secret keys between mobile devices and WiFi back-end services. For the sake of demonstrating our ideas we only focus on the Transmit-SLS phase, noting that they can be extended to other phases of beamsweeping. We next describe Transmit-SLS:\\

\noindent\textit{1)} The initiator (e.g. base station) sends a sequence of beacon frames, one in each sector. The responder (e.g. mobile device) receives these frames with a quasi-omnidirectional antenna pattern. Each beacon frame is marked with an ID for the used antenna array and sector. \\
\noindent\textit{2)} The responder receives the aforementioned frames with varying levels of SNR. It then sends a feedback packet containing the optimal SNR value, and the sector ID of initiator transmitted beacon which was received with this SNR. This feedback packet is transmitted once in every sector of the responder. The initiator receives these frames with a quasi-omnidirectional antenna pattern.\\
\noindent\textit{3)} Upon receiving the feedback packet from the responder, the initiator will be informed of the best sector to use for transmission. The initiator will then send one feedback packet on this sector, containing the optimal SNR value and the ID of the sector used by the responder which was received with this SNR.\\
\noindent\textit{4)} Upon receiving the feedback packet from the initiator, the responder will be informed of the best sector to use for transmission.

\noindent\textbf{System Parameters.} We assume a WiFi network with $N_T$ base stations (which act as Alice's transmitters) and a mobile device with $N_R$ antennas. Each base station is equipped with mmWave planar antenna arrays with $6 \times 6$ elements, while each mobile device is equipped with a single antenna array. The antenna arrays specifications and radiation patterns follow the standard in~\cite{3gppChannel}. As mentioned earlier, we use the channel model proposed in~\cite{thomas20143d} with space consistency as specified in~\cite{3gppChannel}. We assume that $P_T/N_oB = -99 $ dB and the channel bandwidth is $1$ GHz. All transceivers have a noise figure of $-99$ dBm. We assume that base stations have $36$ transmission sectors, with the first sector centered at $0^\circ$ with inter-sector separation of $10^\circ$.

\subsection{Analytical Expressions}
Empirically, and according to the channel model we use, $\gamma^{(i,j)}$ in dB is normally distributed, {\it i.e.}, $f_{\gamma^{(i,j)}} (x) \sim \mathcal{N}(\gamma^{(i,j)}_{\text{av}},\sigma^2)$. The parameters are $\sigma^2 = 24$ and

\begin{equation}
\label{eq::SNR_avdB}
 \gamma^{(i,j)}_{\text{av}}(d,\theta) = \gamma^{(i,j)}_{\text{av}}(1,\theta) - 21 \log_{10}(d),
\end{equation}
where $d$ and $\theta$ are the distance and the azimuth angle between $i$th transmitter and $j$th receiver ($j$ here refers to an antenna of either Bob or Eve), and where we explicate by $\gamma^{(i,j)}_{\text{av}}(d,\theta)$ that $\gamma^{(i,j)}_{\text{av}}$ depends on the factors $d$ and $\theta$.
The term $\gamma^{(i,j)}_{\text{av}}(1,\theta)$ corresponds to the average received SNR at a receiver located $1$ m away from the $i$th transmitter and with the same azimuth angle as the $j$th receiver. This value is dependent on the transmitted power, receiver noise figure as well as the beampattern of the transmitting antenna array. Empirically, $\gamma^{(i,j)}_{\text{av}}(1,\theta)$ can be approximately modeled as

\begin{equation}
\label{eq::SNR_av_dB_theta}
\begin{aligned}
\gamma^{(i,j)}_{\text{av}}(1,\theta) &= G(\theta) + \gamma^{(i,j)}_{\text{init}} 
- 66.8, \\
G(\theta) = 10\log_{10}\left| \left(
\sum\limits_{i=1}^3  0.33 \right.\right. & \left.\left. \cos \left(\dfrac{\theta}{1.8} \right) \cos\left( \dfrac{2i-1}{2} \pi \sin\left( \theta \right) \right) 
\right)^3 \right|, \quad \theta \in [-180,180],
\end{aligned}
\end{equation}
where $\gamma^{(i,j)}_{\text{init}} = 10\log_{10}(\dfrac{P_T}{N_o B})$ and the subtraction of $-66.8$ dB is to account for the path loss due to $1$-m of signal propagation. The empirical expressions discussed here are obtained from Monte-Carlo simulations with $300000$ iterations. Figure~\ref{fig::SNR_av_dB_vs_d} shows $\gamma^{(i,j)}_{\text{av}}$ versus $r$ for $\theta = 0$, both from numerical simulations as well as the expression in~\eqref{eq::SNR_avdB}. Figure~\ref{fig::SNR_av_dB_vs_Theta} shows $\gamma^{(i,j)}_{\text{av}}$ versus $\theta$ for $r = 1$ both from numerical simulations as well as the expression in~\eqref{eq::SNR_av_dB_theta}. Figure~\ref{fig::SNR_hist} shows the empirical histogram of $\gamma^{(i,j)}$ values in dB, as well as the normal distribution $\mathcal{N}(\gamma^{(i,j)}_{\text{av}},24)$ for $r=0.2$ m, $1$ m and $2$ m.

\begin{figure*}
 \centering
 \begin{minipage}{0.49\columnwidth}
  \centering
  \includegraphics[width=\columnwidth]{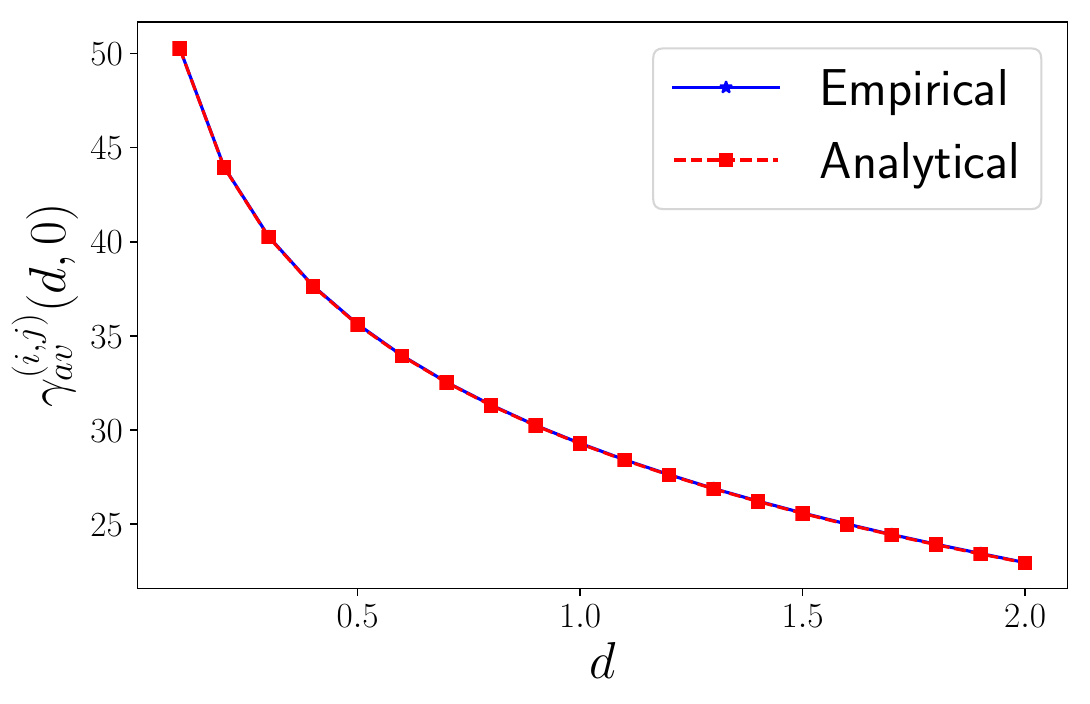}
 \caption{$\gamma^{(i,j)}_{\text{av}}$ versus $r$ for $\theta = 0$.}
 \label{fig::SNR_av_dB_vs_d}
 \end{minipage}
 \begin{minipage}{0.49\columnwidth}
  \centering %
 \includegraphics[width=\columnwidth]{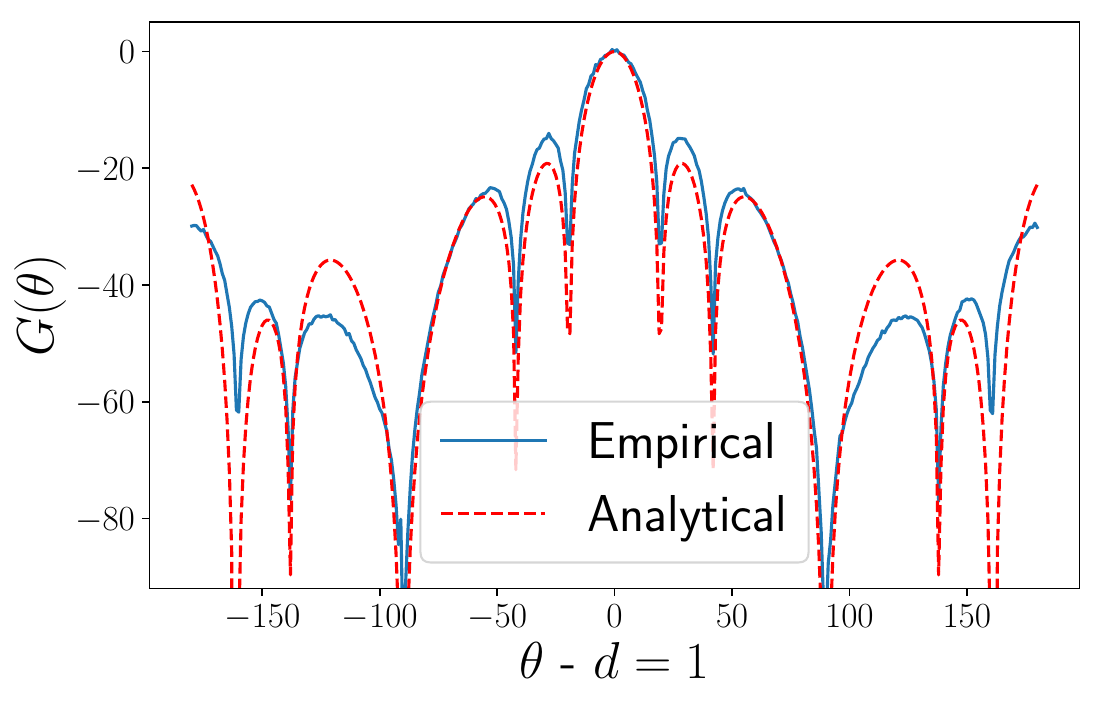}
 \caption{$G(\theta)$ versus $\theta$ for $r = 1$.}
 \label{fig::SNR_av_dB_vs_Theta}
 \end{minipage}
\end{figure*}

\begin{figure}
\centering
\subfigure[$r = 0.2m$]{
\centering
 \includegraphics[width=0.3\columnwidth]{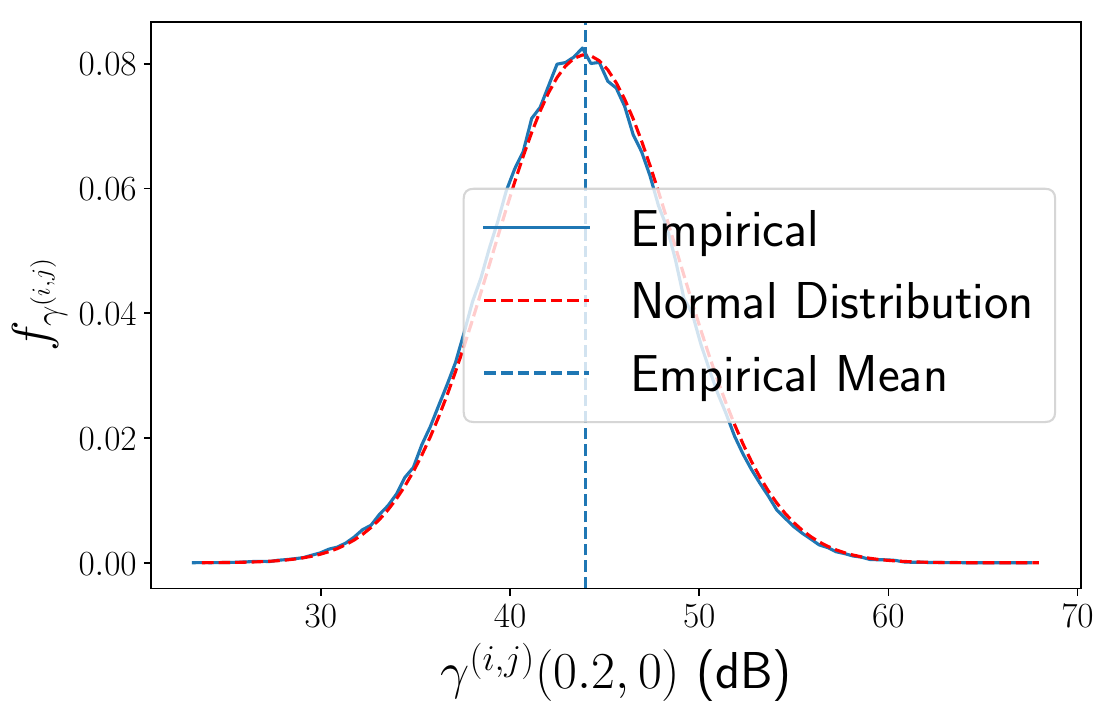}
 \label{fig::SNR_hist_a}
}
\subfigure[$r = 0.6m$]{
\includegraphics[width=0.3\columnwidth]{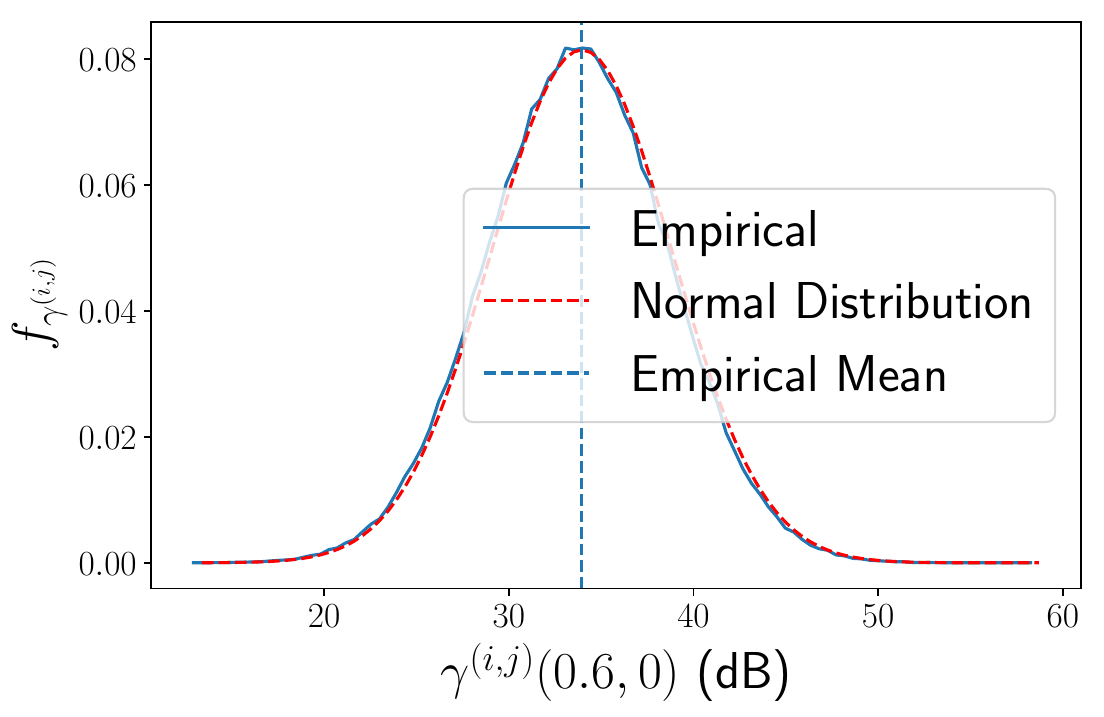}
\label{fig::SNR_hist_b}
}
\subfigure[$r = 1m$]{
\includegraphics[width=0.3\columnwidth]{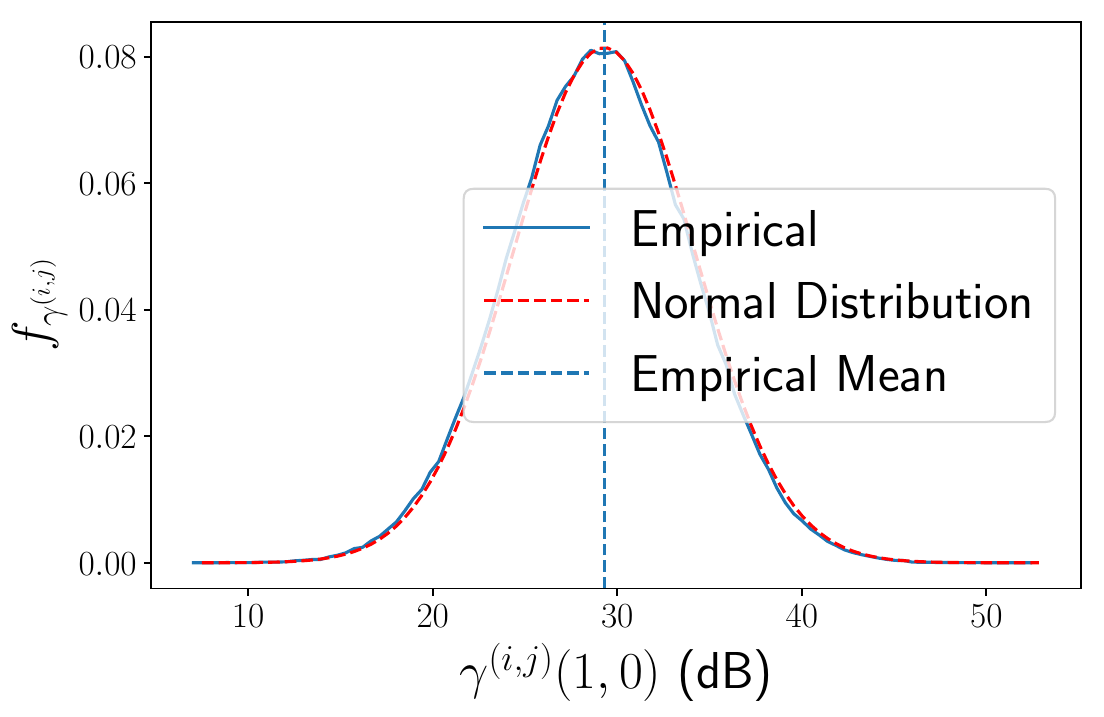}
\label{fig::SNR_hist_c}
}
 \caption{Probability density function for $\gamma^{(i,j)}_{\text{av}}$ - blue line is the empirical distribution and the red line is the analytical expression for the PDF.}
 \label{fig::SNR_hist}
\end{figure}
%
%
%

Based on the aforementioned assumptions, we can express $P_{\text{SecRec}}$ as

\begin{align}
 P_{\text{SecRec}} &= \text{Pr}\left( \exists (i,j) : \gamma^{(t_i,r_j)} \geq \text{Th}_1, \gamma^{(t_i,e_k)} \leq \text{Th}_2 \: ; \: \forall k \right) 
= 1 - \text{Pr} \left(\forall (i,j) : \overline{\gamma^{(t_i,r_j)} \geq \text{Th}_1, \gamma^{(t_i,e_k)} \leq \text{Th}_2 \: ; \: \forall k} \right) \nonumber \\
 &\stackrel{(a)}{=} 1 - \prod\limits_{(i,j)} \text{Pr} \left(\overline{\gamma^{(t_i,r_j)} \geq \text{Th}_1, \gamma^{(t_i,e_k)} \leq \text{Th}_2 \: ; \: \forall k} \right) 
 = 1 - \prod\limits_{(i,j)} \left( 1 - \text{Pr} \left({\gamma^{(t_i,r_j)} \geq \text{Th}_1, \gamma^{(t_i,e_k)} \leq \text{Th}_2 \: ; \: \forall k} \right) \right) \nonumber \\
 &= 1 - \prod\limits_{(i,j)} \left( 1 - Q \left(\dfrac{\text{Th}_1 - \gamma^{(t_i,r_j)}_{\text{av}}  }{\sqrt{24}}\right) \prod\limits_{k} \left(1 - Q\left( \dfrac{\text{Th}_2 - \gamma^{(t_i,e_k)}_{\text{av}} }{\sqrt{24}} \right)  \right)  \right) \nonumber \\
\end{align}
where $(a)$ follows by making an assumption that the variables $\gamma^{(i,j)}, \forall (i,j)$ are independent.

\subsection{Secret Key Establishment Protocols}

Incorporating the secret key exchange scheme into the assumed WiFi network can be done in various ways, each with different levels of effectiveness (in terms of the proposed security metrics) as well as its complexity ({\it e.g.}, how much change in the communication protocol is required to facilitate the scheme). In addition, the performance of the proposed scheme is dictated by the values of $\text{Th}_1$ and $\text{Th}_2$. The thresholds are computed based on the assumed transmission rate and the target secret key rate as follows. Let the beacon frames be transmitted at a rate of $R_T$~Mbps.
Setting the Decoding rate in equation~\eqref{eq::sec_key_gen_rate} to this value would give the corresponding value of $\text{Th}_1$. To achieve a target secret key exchange rate of $R_S$, the Secrecy Overhead rate in equation~\eqref{eq::sec_key_gen_rate} should be equal to $R_T - R_S$; this directly gives the corresponding value of $\text{Th}_2$.\\

We now consider two possible secret key establishment protocols:

\noindent\textbf{1. Beamsweeping-Based (Less Overhead).} In this protocol, the secret key establishment protocol is bootstrapped on top of the T-SLS protocol. Specifically, the protocol includes a chunk of random bits in the beacon frame used by transmitter antennas during the T-SLS phase. 
Beacon frames are transmitted at a rate of $R_T = 27.5$~Mbps~\cite{nitsche2014ieee}.
We assume that a chunk of $1$ kbits in the beacon frame is allocated for random bits. In this chunk, we assume that $250$ random bits are encoded by a wiretap code and inserted. Therefore we have the Secrecy Overhead to be equal to $750/1k \times 27.5$~Mbps. The corresponding values of $\text{Th}_1$ and $\text{Th}_2$ can therefore be computed as described earlier. This approach does not require an intrusive change in the existing transmission protocol of the assumed WiFi network. Therefore, as will be shown in the next section, it allows for the establishment of shared secret keys at virtually no additional transmission overhead. \\

\noindent\textbf{2. Dedicated Secret Key Exchange Packets (High Secrecy Rate).} In this protocol, dedicated frames are sent for the purpose of secret key establishment. The Decoding Rate as well as the Secrecy Overhead are determined so as the establish a good secret key rate and a small insecure area. These dedicated frames are sent after a legitimate receiver is detected by the transmitter at the T-SLS phase, and therefore an estimate of the receiver's SNR value is known by the transmitter.

\subsection{Performance Evaluation}
We assume that a legitimate receiver has $N_R = 1$ one antenna which is located at position $(0,0)$ in space, and receives in an omni-directional way. The transmitter has $N_T$ transmitting antenna array, which are symmetrically distributed around the point $(0,0)$ on a circle with radius $d$. 
We also assume that the eavesdropper is equipped with one antenna which receives in an omni-directional manner (similar to the receiver). 
When a transmitter antenna array is transmitting data to the legitimate receiver antenna, we assume that the transmitter antenna array beamforms at the location of the receiver's antenna, even during T-SLS phase ({\it i.e.}, we assume that the legitimate receiver is always aligned with the main direction of a sector).
This assumption is reasonable given the locations of transmitters we consider with respect to the legitimate receiver. 

\begin{figure*}
\centering
\begin{minipage}{0.45\columnwidth}
 \centering
 \includegraphics[width=\columnwidth]{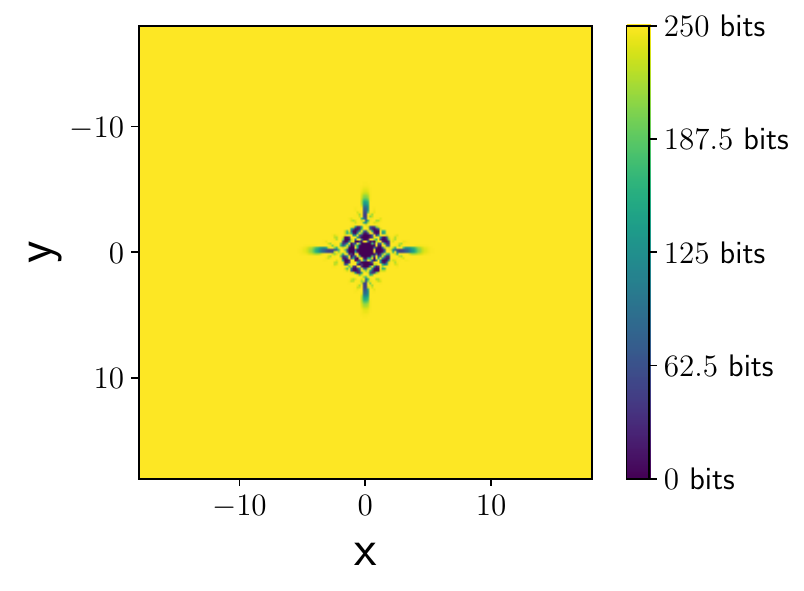}
 \caption{Beamsweeping-Based - $R_{\text{av}}$ for $d = 2$, $N_T = 4$ and $\delta = 0.01$.}
 \label{fig::R_av}
\end{minipage}
\begin{minipage}{0.45\columnwidth}
 \centering
 \includegraphics[width=\columnwidth]{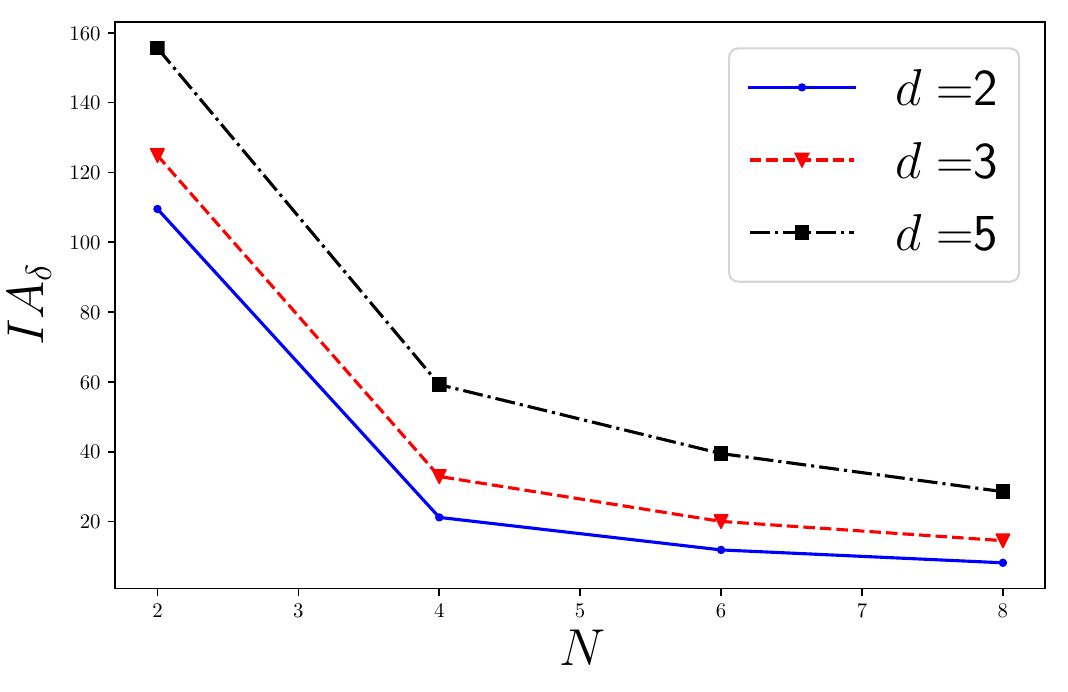}
 \caption{Beamsweeping-Based - $IA_{\delta}$ versus $N$ for $d = 2, \: 3, \: 5$ and $\delta = 0.01$.}
 \label{fig::IA_vs_N}
\end{minipage}
\end{figure*}

\begin{figure*}
 \centering
 \begin{minipage}{0.45\columnwidth}
 \centering
 \includegraphics[width=\columnwidth]{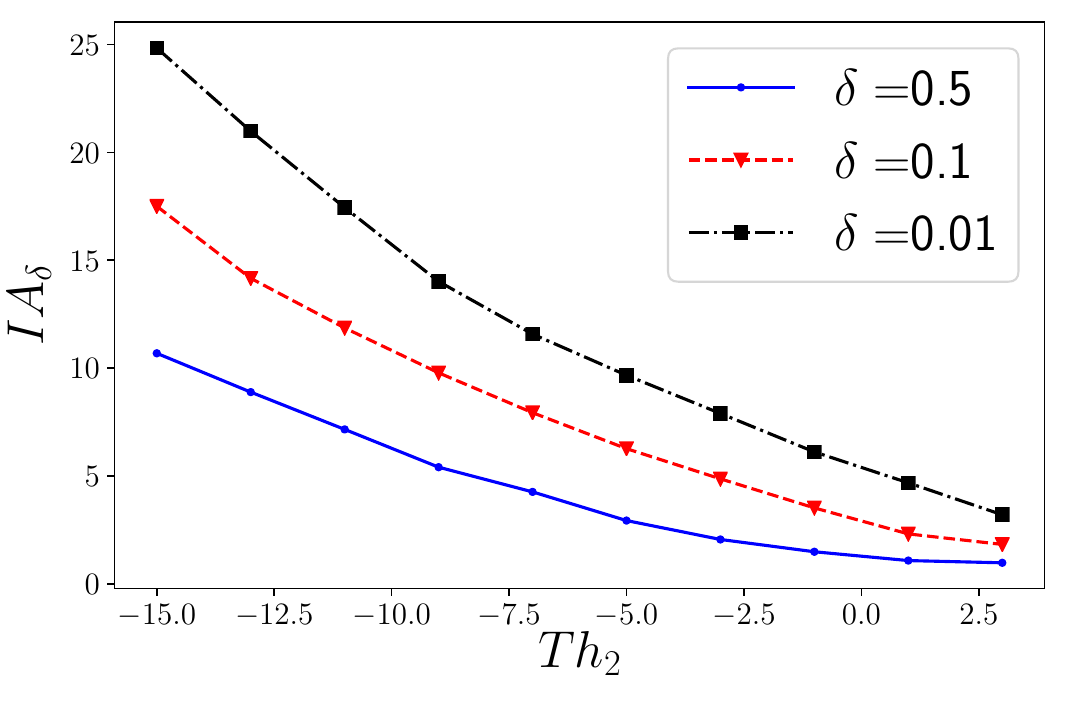}
 \caption{$IA_{\delta}$ versus $\text{Th}_2$ for $\delta = 0.01, \: 0.1, \: 0.5$, Th$_1 = 7$ dB and $d = 2$.}
 \label{fig::IA_vs_Th2}
\end{minipage}
 \begin{minipage}{0.45\columnwidth}
  \centering
 \includegraphics[width=\columnwidth]{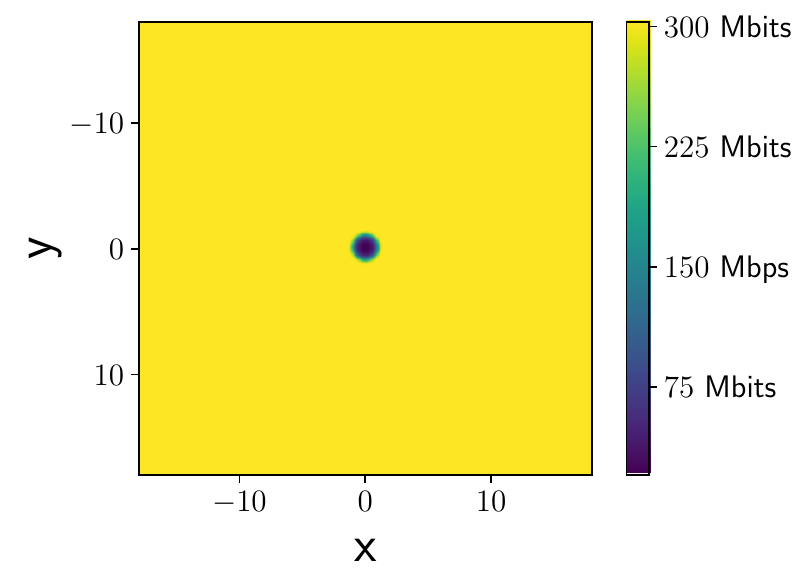}
 \caption{$R_{\text{av}}$ for $\text{Th}_1=7$, $\text{Th}_2 = 3$, $\delta = 0.01$ and $d = 2$.}
 \label{fig::R_av_good}
 \end{minipage}

\end{figure*}

\textbf{1) Beamsweeping-Based.} Figure~\ref{fig::R_av} shows the average number of secret bits when $N_T = 4$ and assuming a Beamsweeping-Based protocol. The value of $R_{\text{av}}$ depends on the probability of Secret Reception, and therefore is dependent on the location of the eavesdropper. Nevertheless, Figure~\ref{fig::R_av} shows that an average of $250$ bits of secret keys can be achieved against eavesdroppers $5$ meters away from the legitimate receiver antenna. Therefore, our simulations suggest T-SLS can automatically provide an average of $250$ bits between the transmitter and receiver which are kept secret from eavesdroppers that are $5$ meters away from the receiver. 

Figure~\ref{fig::IA_vs_N} shows the insecure area $IA_{\delta}$ versus $N_T$ for different values of $d$. We consider here $\delta = 0.01$. For $N_T = 4$ and $d = 2$, the insecure region ({\it i.e.}, the area in which an eavesdroppers renders $P_{\text{SecRec}} \leq 0.99$) is approximately $20$ m$^2$. As intuitively expected, increasing $d$ further increases the insecure area. Moreover, Figure~\ref{fig::IA_vs_N} suggests that further increasing the number of transmitters does not always decrease the insecure area -- the amount of decrease in the insecure area as $N_T$ increases diminishes, with a seeming plateau reached for the case of $d = 2$ at approximately $N_T = 8$.

\textbf{2) Dedicated Secret Key Exchange Packets.}
The relatively high values of an insecure areas exhibited in Figure~\ref{fig::IA_vs_N} is due to the particular choices of $\text{Th}_1$ and $\text{Th}_2$ so as to be compatible with T-SLS frames. On the other hand, if dedicated frames (with specific Decoding Rates and Secrecy Overhead) are to be used, better-performing (in terms of average secret key exchange rate and insecure areas) secret key exchange protocols can be established. Figure~\ref{fig::IA_vs_Th2} shows the insecure area versus different choices of $\text{Th}_2$ for $d = 2$, $N_T = 4$ and $\text{Th}_1 = 7$. The figure suggests that the insecure region can be significantly decreased (approximately to an order of magnitude) by increasing the value of $\text{Th}_2$ to $3$ dB. In fact, using these particular choices of the thresholds yields a maximum secret key generation rate of $301$ Mbps. Figure~\ref{fig::R_av_good} shows how much the average secret key rate is achieved against eavesdroppers in different locations. It is clear that in most of the region, the maximum secret key generation rate is achieved.

\section{Showcase II - Vehicle platooning}
\label{Sec:platoon}

Vehicle platooning comprises a set of autonomous cars which drive on the road in a line formation with approximately the same speed and relatively small inter-vehicle distances \cite{ploeg2011design}. 

\noindent\textbf{Setup and Protocol.} 
We assume that each car has two mmWave antenna arrays used for transmission and two omni-directional antennas for reception. One pair of transmit/receive antennas (pair-1) is mounted on top of the roof of the car at a height of 0.5 m and the other pair (pair-2) at 1 m. 
We assume that $N_o = -80$ dBm and $P_T = 30$ dBm.

The secret key protocol 
works as follows: 1) Pair-1 from the front car sends random packets encoded with a suitable wiretap code to Pair-1 of the back car (Link-1), 2) Pair-2 from the front car sends random packets encoded with a suitable wiretap code to Pair-2 of the back car (Link-2), 3) the front car sends a set of carefully-designed packets as per the protocol in \cite{czap2015secret} to the back car to establish secret keys.

\noindent\textbf{Analysis and Discussion.} The preliminary channel model we developed does not account for random channel fading. Therefore, the concept of $\delta$-insecure area becomes a deterministic one, {\it i.e.}, we only consider $IA_{0}$. Our key agreement protocol can establish up to $166$ Mbps of secret bits, with $IA_{0} = 0$. 
To put this number in perspective, a typical symmetric key exchange algorithm such as (DH-2048), implemented on an off-the-shelf Dedicated-Short-Range-Communication (DSCR) transceiver, gives a key generation rate of approximately $20$ kbps; that is, there is a performance gain of approximately $4$ orders of magnitude. Table~\ref{table::comp} shows a comparison between DH-2048 and our proposed secret key establishment protocol.
%


\noindent\textbf{Application Example.} We {will} show next that, thanks to the high rate of secret key generation, our protocol allows for the use of OTP to secure the string stability functionality of vehicle platoons.
In order to maintain string stability within the platoon controllers, each car in the platoon exchanges data packets every 100 ms, each of size $60$ bytes~\cite{ploeg2011design}, with both the cars in front of and to the back of it.
We will show that our suggested key agreement technique can generate enough secret bits which allows the use of OTP to encrypt such messages.
Assume that our proposed algorithm for key generation is used every $5$ minutes for a duration of $10$ ms. Therefore, each two consecutive cars will have an amount of secret keys equal to $10$ ms $\times 166$ Mbps $\approx 200$ kB to use for encryption during the next $5$ minutes. 
The total amount of data to be transmitted during the next $5$ minutes is equal to $5$ min $\times$ $60$ B $\approx 180$ kB $\leq 200$ kB of secret bits. Therefore, {OTP is a practical solution, something rarely achieved in any other kind of security application.}

\begin{table}
 \centering
 \begin{tabular}{|c|c|c|}
 \hline
  & DH-2048 & Erasure-based mechanism \\
  \hline
    Critical resource & Computation power & Bandwidth \\
    \hline
  \begin{tabular}{l}Secret Key Rate\\(realistic setup)\end{tabular} & $20$ kbps & $166$ Mbps \\
  \hline
  \begin{tabular}{l}Complexity of\\encryption technique\end{tabular} & Moderate (AES) & Simple (OTP) \\
  \hline
  Quantum-Vulnerable & Yes & No (Info. theoretically secure) \\
  \hline
  \begin{tabular}{l}Adversary with\\high network\\ presence\end{tabular} & Resilient & Weak \\
  \hline
 \end{tabular}
\caption{Comparison between DH and proposed mechanism for vehicle platooning.}
\label{table::comp}
\end{table}

\noindent\textbf{Discussion.} Comparing our proposed key establishment mechanism that is based on channel erasures, against conventional DH algorithms, we note that our solution is superior in the following aspects: 1) it attains $4$ orders of magnitude gains in terms of key generation rates, 2) it allows for using encryption techniques with very low complexity (e.g., OTP) and 3) it is not vulnerable against eavesdroppers with high computational powers (e.g., quantum adversaries). However, it is affected by the availability of a wide transmission bandwidth and the network-presence of adversaries.

\section{Conclusion and Discussions}
\label{Sec:concl}
In this paper we investigated how the directional nature of mmWave communication can be used to enhance security. 
We showcased how mmWaves and wiretap codes can enhance the performance of secret key generation techniques in the context of two applications, over-$60$ GHz-WiFi networks and vehicle platooning. For both cases, we used/developed channel models with realistic antenna parameters to give realistic assessment of such protocols. For the case of WiFi networks, we empirically developed analytical expressions for the received SNR.
We showed that 
existing T-SLS protocol in IEEE 802.11ay can be used to create a few hundred secret keys at virtually no additional transmission cost, while 
dedicated protocols in both cases establish very high rates. This work is an initial investigation on the topic. We believe that our results are enticing enough to build a complete system-level implementations of our proposed scheme and analyze its performance in real-world.

\bibliographystyle{IEEEtran}
\bibliography{Bib}

\end{document}